\documentclass[11pt,twoside]{article}
\usepackage{asp2010}
\resetcounters

\newcommand{\lsp}{LS~I~+61$^{\circ}$303}
\newcommand{\lsi}{LS~I~+61$^{\circ}$303~}

\begin{document}


\title{Microblazar properties of \lsp}

\author{Maria Massi
\affil{MPIfR, Auf dem H\"ugel 69, D-53121 Bonn, Germany}
}
\author{Eduardo Ros
\affil{Dep. Astronomia, Univ. de Val\`encia, E-46100 Burjassot,
Valencia, Spain}
\affil{
MPIfR, Auf dem H\"ugel 69, D-53121 Bonn, Germany}
}
\author{Lisa Zimmermann
\affil{MPIfR, Auf dem H\"ugel 69, D-53121 Bonn, Germany}
}
\author{Guidetta Torricelli-Ciamponi
\affil{INAF, Oss. Astrofisico di Arcetri, Largo E. Fermi 5,
I-50125 Firenze, Italy}
}

\begin{abstract}In this paper we describe  the gamma-ray binary \lsi as a blazar like object, based on 
variable Doppler boosting and core-shift effect considerations.
\end{abstract}


\section{Introduction}   
\label{introduction}
In Active Galactic Nuclei (AGN),
a super-massive black hole 
of $10^6-10^9$  solar masses
accretes matter from its host galaxy.
An important subclass of the AGN are blazars,   AGN whose  radio-emitting  jets form a small angle with respect to the line of sight.
The class of the  X-ray binaries  is very similar to the AGN class
with respect to  the accretion/ejection physical process.
In X-ray binary systems  the ``energy-engine''
  is  a compact object of only a few
  solar masses accreting from the companion star.
  If an  X-ray binary shows evidence for a radio-jet, 
  it is referred to as a microquasar;
 a microquasar whose jet points towards us would be a microblazar
  \citep*{mirabelrodriguez99, kaufman02}.

\section{Doppler boosting}

Doppler boosting is an effect predicted by the theory of special relativity
that enhances the radiation
from material that is moving towards the observer (i.e., small viewing angle) at nearly the speed of light,
and attenuates it, if it moves in the opposite direction at such a speed.
Doppler boosting strongly depends on the angle between the ejecta and the observer's line of sight.
A precession of the jet implies a variation of this angle and therefore
variable Doppler boosting.
In \lsi we can observe position angle variations of the jet and  flux attenuation of the receding jet.
At some epochs the receding jet is  attenuated at a level under the sensitivity of the radio images
and then only the approaching jet is visible,  i.e., the jet appears with the  one-sided structure typical of blazars.
The ten  consecutive VLBA  images, interlapsed 3 days and labeled from A to J,  presented in
\citet*{massi12}  show  different position angles
and changes from a one-sided to a double-sided structure. 
 The first image (A) shows a double-component structure, the second image
 (B) a  double twisted jet as expected for a fast precession,  some  images
 (C, F, I, and J) show a one-sided jet and  others (D, E, G,  and H)  a  double-sided jet.
  The strong attenuation of the receding jet indicates a rather small ejection angle in \lsp, but
  also probes that the velocity of the jet is relativistic otherwise there would be no Doppler deboosting 
  of the receding jet  and the two jets would always be visible. 

\section{Core-shift  effect in the jet of \lsp}
The millimeter-wave core in AGN
is a physical feature of the jet, i.e., coincident with
the  standing shock formed at few Schwarzschild radii from the engine  \citep{jorstad10}.
Indeed, recent  six frequency observation of  M87  proved that the
7~mm-core ($\nu$=43 GHz) is located  within only 23 Schwarzschild radii
   from the system center \citep{hada11}.
     This  is  different from the so called  cm-wave core, the most compact feature in the jet image,
     the location of which is determined by jet opacity.
In particular, with the cm-core  being the surface where
the optical depth  is equal to unity, its location  is a function of observing wavelength,
particle density, magnetic field, and jet geometry.
In microquasars,  the analoguos standing shock should be   the near infrared-core (300 THz)  
where the turnover from optically thick to optically thin emission occurs \citep{russell06}.   
Therefore, the cm-core is very  displaced from the system center. 
In \lsp, one can perform three different comparisons. First, one can compare  \citep[Fig. 3]{massi12}
the positions of the 
13 cm-core with those of the 3.6 cm-core. 
Indeed, the astrometry  at 13 cm,  even if strongly affected by large Galactic scattering and
resolution problems, shows   
a global  offset.
The second one compares  the different positions of the 3.6~cm-core in the consecutive VLBA observations
and shows
that the  peaks trace well an ellipse with a semimajor axis of 1.35 mas.
Since the jet is precessing around a cone of unkown aperture $\psi$, 
one can assume  that the ellipse is  the cross-section of the
precession cone at the distance $L\,\cos\,\psi$ of the 3.6 cm-core, with
$L\, \sin\,\psi\,  \cos i$=1.35 mas, where $i$  is the inclination of the orbit.
The last one compares the flux density, $S$, of the different images at  3.6 cm. 
Whereas all images but C have   $S=18-42$ mJy the flux density for C
is 95$\pm$8 mJy. Nevertheless,  C remains on the ellipse. This implys that  the eventual change, 
 $\Delta L$, associated to
the $\Delta S$ must remain  within the astrometry error of 0.04 mas (Massi et al. in preparation).



\end{document}